\documentclass[twocolumn,showpacs,preprintnumbers,amsmath,amssymb]{revtex4}

\usepackage{graphicx}
\usepackage{dcolumn}
\usepackage{bm}

\begin{document}

\title{Fermionic atoms trapped in 
one-dimensional optical superlattice with harmonic confinement}

\author{
Takanori Yamashita and Norio  Kawakami
}

\affiliation{
Department of Applied Physics, Osaka University, Suita, Osaka 565-0871, Japan}

\author{Makoto  Yamashita}
\affiliation{
NTT Basic Research Laboratories, NTT Corporation, 3-1, Morinosato-Wakamiya, 
Atsugi-shi, Kanagawa 243-0198, Japan
}%

\date{\today}
\begin{abstract}
We study the ground-state properties of spin-1/2 fermionic atoms confined 
in a one-dimensional optical superlattice with harmonic confinement by using 
the density matrix renormalization group method. For this purpose, we consider 
an ionic Hubbard model that has superlattice potentials with 2-site periodicity. 
We find that several different types of insulating regimes coexist 
even if the number of atoms at each site is not an integer, but its average 
within the unit cell is an integer or half integer. 
This is contrasted to the coexisting phase of the metallic and Mott-insulating 
regimes known for the ordinary Hubbard model in an optical lattice. 
The phase characteristics are elucidated by investigating the profiles of the 
atom density, the local density/spin fluctuations, the double occupation probability 
and the spin correlations in detail.
\end{abstract}

\pacs{03.75.Ss, 05.30.Fk, 34.50.-s, 71.30.+h}

\maketitle

\section{Introduction}\label{sec:int}
Experimental techniques for manipulating ultracold atoms have made great 
progress since the successful realization of atomic Bose-Einstein condensates 
(BECs) \cite{ColdAtom}. 
Optical lattices, formed by a standing wave of laser light, are providing the ideal 
stages for an experimental investigation of the fundamental many-body 
problems in condensed matter 
physics via ultracold atomic gases \cite{Bloch, Bloch2, Jaksch, Morsch}. 
A recent series of experimental studies on bosonic Mott-insulators 
\cite{Greiner, Mandel, Stoferle, Paredes, Folling, Gerbier, Gerbier2} 
have clearly demonstrated this feature, 
namely the realization of quantum simulators. 
\par
A noteworthy advantage of atomic gases over condensed matter such as solids or 
liquids is that their experimental parameters are highly controllable. 
Both density and temperature of atomic gases are fully controlled in the process of 
evaporative cooling \cite{ColdAtom}. The depth of the optical lattice potential is controlled 
by the light intensity, which leads to the precise alternation of atom tunneling 
between adjacent sites \cite{Bloch,Bloch2}. Furthermore, we can create one-, two-, and 
three-dimensional optical lattices depending on the laser beam configuration with the use 
of interference effects \cite{Bloch,Bloch2}. 
Even the interactions between the atoms can be widely tuned with the help of 
Feshbach resonances \cite{Inouye}. Therefore, in the experiments, it is possible to precisely 
change the important system parameters such as density, temperature, tunneling, 
dimensionality of lattices, interaction strength and so forth.  Ultracold atoms in optical 
lattices allow us to investigate the quantum many-body effects in a well-controlled manner.
\par
Experiments with quantum degenerate fermionic gases have already clarified the 
intriguing physics of such phenomena as a molecular 
BEC \cite{Greiner2, Jochim, Zwierlein} and 
the crossover between a BEC and a Bardeen-Cooper-Schrieffer (BCS) superfluid 
\cite{Regal, Zwierlein2, Bartenstein, Bourdel}. 
Ultracold fermionic atoms trapped in optical lattices are now 
stimulating both experimental and theoretical interest. 
This system is well captured by the Hubbard model Hamiltonian \cite{Rigol}, 
which is widely discussed with respect to strongly correlated electron systems. 
Several theoretical analyses have predicted the possibility of 
the Mott metal-insulator transition (MMIT) in this system 
\cite{Rigol, Rigol2, Machida, Liu, Campo, Xianlong}. 
Although experimental investigations are still under way 
\cite{Modugno, Ott, Kohl, Stoferle2}, 
the recent demonstration of a fermionic band insulator in three-dimensional 
optical lattices \cite{Kohl} has convinced us that MMIT will be observed 
in near future. A precise comparison of the theory and experiments  
will provide us with a deeper understanding of the quantum phase 
transition in fermionic systems.  
\par
The {\it superlattice} geometry of optical lattices, 
i.e., optical superlattice, was successfully demonstrated 
as an advanced manipulation technique in a recent report 
\cite{Peil}. This technique has been introduced to control the distribution of atoms on 
lattices more functionally and will play a key role in the implementation of 
quantum computing with ultracold atoms \cite{Peil, Jaksch}.
On the other hand, in condensed matter physics, it is known that lattice distortions 
generate a variety of many-body effects in strongly correlated electron 
systems 
such as the charge-transfer organic materials and ferroelectric perovskites 
\cite{Hubbard,Nagaosa,Egami,Ishihara,b8,b9,Fabrizio,b12,b16,b17,b18,b19,b20,Manmana, Otsuka}. 
The ultracold fermionic atoms in optical superlattices \cite{Paredes2} 
can be regarded as a quantum simulator for these materials. 
Our present study is highly motivated by these important potentialities of 
optical superlattices. 
\par
In this paper, we investigate the ground state properties of 
S=1/2 fermions trapped in a one-dimensional optical 
superlattice with 2-site periodicity. 
The external harmonic confinement inherent in the real experiments \cite{Bloch} 
is further considered in our model. 
We show that a new coexistence phase appears consisting of  
three distinct insulating regions caused by the 
superlattice potentials, which is 
in contrast to the ordinary Hubbard model in the optical lattice.
This is elucidated via a detailed analysis of the 
local density of atoms and spin correlations by using the density matrix 
renormalization group method.
Furthermore, we find unique spin correlations between
the insulating regions when these regions are separated by a 
metallic region. 
\par
This paper is organized as follows. In the next section we 
briefly mention the model and the method. We then show 
the density profiles of fermions and several 
density/spin correlation functions to clarify the zero temperature properties.
The final section provides a brief summary. 
\par

\section{Model and Method}\label{sec:met}\label{sec:mod}

\begin{figure}[t]
\begin{center}
\includegraphics[width=.75\linewidth]{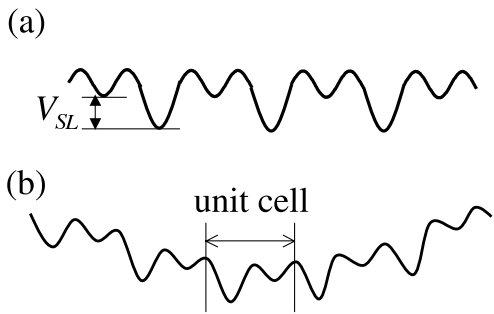}
\caption{Potential profiles of (a) 2-site periodic superlattice 
 and (b) with a confining potential. 
 $V_{SL}$ is the difference between the potential energies of odd 
and even sites. 
}
\label{fig:pp}
\end{center}
\end{figure}

We consider a gas of fermionic atoms embedded in an optical superlattice 
 with an alternating superlattice potential and a confining 
 parabolic potential 
 (see Fig.\ 1(b)), which is described by the Hubbard Hamiltonian, 
\begin{eqnarray}
{\cal H} &=&  -t \sum_{i,\sigma} (c_{i,\sigma}^{\dagger} 
c_{i+1,\sigma} +h.c.)
+ U \sum_i n_{i \uparrow}n_{i \downarrow} \nonumber\\
& & \!\!\!\!\! +\sum_{i,\sigma} V_i n_{i \sigma}
+ V_c \sum_{i\sigma}(i-{N_s}/2)^2 n_{i\sigma}
\label{eq:model}
\end{eqnarray}
 where $c^\dag_{i\sigma}$, $c_{i\sigma}$ and 
 $n_{i\sigma} = c^\dag_{i\sigma}
 c_{i\sigma}$ are respectively the fermion creation, annihilation and
number operators relevant to the site labeled $i$ with spin
 $\sigma(=\uparrow, \downarrow)$. Here, $U$ is the on-site repulsion ($U>0$),
 $V_c$ the curvature of the parabolic confining potential, 
$V_i$ the local potential at site $i$ and $t$ the hopping 
amplitude between  nearest neighbor sites. 
 
We define the local potential $V_i$ with 2-site periodicity as 
\begin{eqnarray}
    {V_i} = V_{SL}\{\bmod(i+1,2)\}
\label{eq:sl}
\end{eqnarray}
where $V_{SL}$ is the difference between the potential energies of 
odd and even sites (see Fig.\ 1(a)). We denote the total number of 
fermions (lattice sites) as $N_f$ ($N_s$).


There are several kinds of numerical methods that can be employed with our fermionic 
superlattice Hubbard model. The numerical diagonalization method can be applied 
straightforwardly to fermionic models, but it restricts our analysis to rather small 
systems with a typical lattice size of 20 sites, which is not large enough for our purpose. 
The quantum Monte Carlo simulations method is an efficient way to deal with quantum 
systems at finite temperatures, and this approach has been applied successfully 
to bosonic Hubbard models to elucidate 
the transition between the superfluid and Mott insulating phases in optical lattices
\cite{Batrouni,Kashurnikov,Wessel}. 
This method, however, suffers from negative-sign problems in fermionic cases, 
which makes it difficult to obtain reliable results at low temperatures. 
The density matrix renormalization group (DMRG)\cite{White,Scollwock} is another powerful numerical 
method with which we investigate quantum systems such as fermionic models with extremely high accuracy, 
although it is restricted to one-dimensional cases. This technique allows us to deal with 
a very large system as required for the analysis of optical lattices. 
In this paper we therefore exploit a variant of the DMRG 
(the so-called finite-system DMRG) \cite{White} 
to study our fermionic Hubbard model in an optical superlattice at zero temperature. 
The essence of the idea is that we first diagonalize a given small system, 
and then enlarge the system size step by step via a renormalization procedure 
by retaining relevant low-energy states which have large eigenvalues of the density matrix. 
This simple procedure is known to give precise results for one-dimensional quantum systems. 
This method has already been reported in detail \cite{White,Scollwock}. 
In our nonuniform model with harmonic confinement, the numerical calculation can be done in parallel 
to the ordinary case without confinement. The only point we have to pay attention to is that 
the superlattice potential should be properly included in each renormalization step.
\par
We restrict our discussions to the singlet ground state with an even number of fermions, 
which are tractable by the DMRG method \cite{White,Scollwock}. 
As far as the local density and its variance are concerned (see Sec.III A and B), 
the even-number case is sufficient to discuss the characteristic properties. 
A relevant difference between the even and odd cases may appear in the spin correlations (Sec.III C) 
since a free S=1/2 spin remains unscreened in the system for the odd case. 
The free spin is not important physically since it does not affect the global nature of the system, 
but rather acts as a magnetic impurity.
\par
\section{Results}

\subsection{Density profiles}

\begin{figure}[t]
\includegraphics[width=0.65\linewidth]{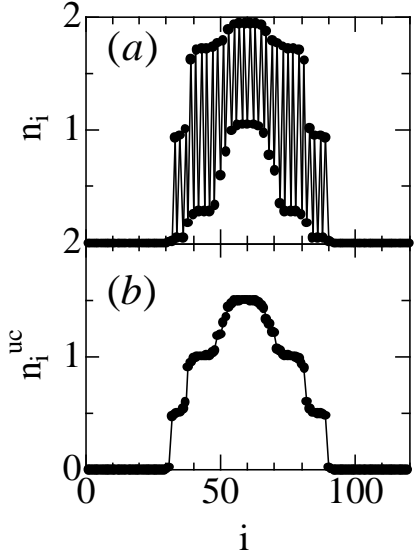}
\caption{(a) Density profiles 
for a trap with $N_s = 120$, $N_f = 60$, 
$V_c/t = 0.01$, $ U/t = 4.0$ and 
 $V_{SL}/t = 6.0$. The filled circles denote the local density $n_i$ and 
 the solid lines are a guide for the eye.
(b) shows the density profile,  $n_i^{uc}$, averaged
within the unit cell. There are plateaus at 
$n_i^{uc}=0.5,1.0,$ and 1.5.
}
\label{fig:local}
\end{figure}

In our optical lattice, the spatial distribution of atoms is 
inhomogeneous because of the confining harmonic potential. We thus 
consider the profile of the local density, i.e. 
the spatial distribution of $n_i$, to be
\begin{equation}
n_i \equiv \sum_{\sigma}
\langle c_{i \sigma}^{\dagger}c_{i \sigma} \rangle.
\end{equation}
%
%
In Fig.\ \ref{fig:local}(a), we show the density profile for
 $N_f=60$ fermions trapped in a lattice 
with the linear size $N_s=120$,  where the other parameters are 
$V_{SL}/t=6.0$, $U/t=4.0$ and $V_c/t=0.01$.
It is seen in Fig.\ \ref{fig:local}(a) that the amplitude of the local density 
alternates between  even and odd sites reflecting the 2-site 
periodic  potential. When there is no superlattice structure, 
namely $V_{SL}$=0, it is 
known that a Mott insulating region appears characterized by 
plateau formation at $n_i=1.0$ \cite{Rigol, Rigol2}.
In the present case, however, the density profile reveals 
more complicated sawtoothed structures, from which we can 
not immediately distinguish metallic and insulating regions.
 To characterize the insulating region clearly,
we therefore introduce the average number of atoms in the unit cell as,
\begin{equation}
n_i^{uc} \equiv 
\frac{1}{2}( n_i +  n_{i+1}).
\end{equation}
In Fig.\ \ref{fig:local}(b), we show the $n_i^{uc}$ value obtained by averaging 
the data of (a) in the unit cell. Plateaus (or shoulders) appear 
at $n_i^{uc}$ =0.5, 1.0, and 1.5, which characterize three distinct 
insulating regions.

\begin{figure}[h]
\includegraphics[width=1.0\linewidth]{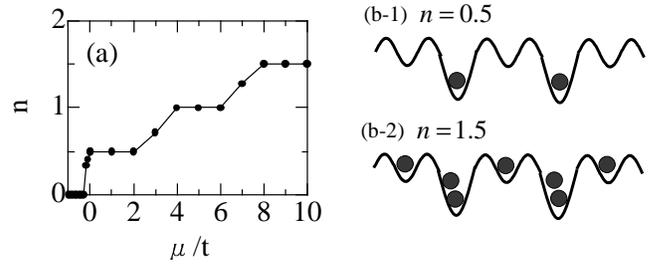}
\caption{Metallic and insulating  phases in the 
one-dimensional homogeneous superlattice system without 
harmonic confinement for $V_{SL}/t=6.0$ and $U/t=4.0$:
(a) density of atoms, $n$, 
as a function of the chemical potential $\mu / t$.
The plateaus, which are formed clearly at 
$n$=0.5, 1.0, and 1.5, imply the existence of well-defined 
insulating phases. (b-1) and (b-2) are schematic diagrams of 
the insulating phases for $n=0.5$ and $1.5$.
}
\label{fig:dp}
\end{figure}

To understand the above characteristic behavior in $n_i^{uc}$,  
it is instructive to observe the bulk properties of the system without
harmonic confinement, which are shown in Fig.\ 3. 
Figure\ \ref{fig:dp}(a) shows the infinite 
DMRG results for filling factor
$n$ as a function of chemical potential $\mu$. 
We can see the formation of 
well-defined plateaus at $n=$0.5, 1.0, and 1.5, which clearly feature the insulating 
phases separated by the metallic phases. The insulating phases  with
$n=$0.5 and 1.5 are caused by the superlattice potential with 
2-site periodicity. The corresponding spatial distribution of atoms 
is schematically illustrated in Fig.\ \ref{fig:dp}(b).
When $n=0.5$, as shown in Fig.\ \ref{fig:dp}(b-1), 
the system is in a variant of
the Mott insulating phase where each site with lower (higher) potential $V_i$
accommodates one (no) atom. On the other hand, 
when $n=1.5$, as shown in Fig.\ \ref{fig:dp}(b-2), the system is 
another Mott type insulator where each site with higher potential 
accommodates one atom, while that with lower potential is almost fully
occupied by two atoms.  
The insulating phase with $n=1.0$ changes its nature depending 
on the interaction strength $U$ in a slightly complicated way, 
from the band insulator (small $U$) to the Mott insulator 
(large $U$). Actually this problem has been extensively 
discussed so far, which is briefly summarized here. Early 
studies by the numerical \cite{b8} and weak-coupling bosonization 
methods \cite{b9} suggested that a single phase transition occurs 
from the band insulator to the Mott insulator when $U$ is 
varied at half filling. On the other hand, Fabrizio {\it et al.}
 predicted a new intermediate insulating phase ``so-called bond-charge 
density wave (BCDW) phase" between them \cite{Fabrizio}. Since their 
proposal, extensive investigations have been done to confirm it. 
Quantum Monte Carlo simulations \cite{b12} supported the existence 
of the intermediate phase. By applying the method of topological 
transitions in spin and charge Berry phases, Torio {\it et al.} 
presented a ground-state phase diagram, which is consistent with 
the scenario of Fabrizio {\it et al.} \cite{b16} The DMRG 
calculations \cite{b17,b18,b19,b20} have been performed by 
several groups. For example, the calculation of various 
structure factors supported the existence of intermediate 
phase \cite{b19} though Kampf {\it et al.} \cite{b20} 
were not able to resolve two transitions clearly.  
Recent calculations based on the DMRG \cite{Manmana} 
and the level-spectroscopy method \cite{Otsuka} with 
high accuracy  have confirmed the existence of the intermediate BCDW phase.

The above results for the 
homogeneous system show that the plateau regions at
$n_i^{uc}$ =0.5, 1.0, and 1.5 in the system with confining potential
have the above insulating properties, which are sandwiched by
 metallic regions. 

\begin{figure}[t]
\includegraphics[width=0.8\linewidth]{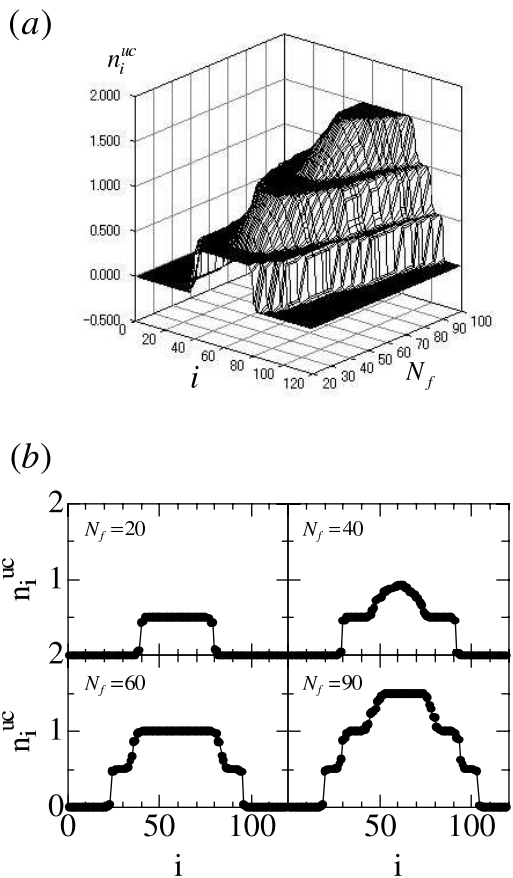}
\caption{(a) Profiles of the average density, $n_i^{uc}$, along the 
trap for different fillings. The flat terraces are the insulating regions.
The parameters are $N_s = 120$, $U/t = 4.0$, $V_{SL}/t = 6.0$,
and $V_c/t = 0.005$. In (a), the number of atoms is 
changed from 20 to 104, while in (b) the number is fixed as
 $N_f$ =20, 40, 60, and 90.
}
\label{average}
\end{figure}

Let us now observe how the density profile of the system changes 
when the atom number $N_f$ is changed systematically.
In Fig.\ \ref{average}(a), we show the average density of atoms, 
$n_i^{uc}$, when $N_f$ is changed continuously.  It is seen that the system 
enters several distinct ground states as $N_f$ increases. 
 In Fig.\ \ref{average}(b), for reference,  we show 
cross-sectional views of (a) 
 by selecting several typical $N_f$ values.
For $N_f=20$, there is an insulating region with $n_i^{uc}=0.5$ in the middle 
of the system.  Just bedside this region, there are 
narrow metallic region windows
with $n_i^{uc}<0.5$, which smoothly continue to the
vacant-atom region $n_i^{uc} \sim 0$.  The system is therefore in a 
phase where metals and insulators coexist, as is the case for an ordinary 
optical lattice without superlattice potentials \cite{Rigol,Rigol2}.
For $N_f=40$, a metallic region with 
$0.5<n_i^{uc}<1.0$ appears in the middle of the system, so that the 
insulating regions with $n_i^{uc}=0.5$ 
are now sandwiched by the two metallic regions with 
$n_i^{uc}<0.5$ and $0.5<n_i^{uc}<1.0$. For 
$N_f=60$, another insulating phase with $n_i^{uc}=1.0$ 
emerges in the middle region, and 
a further increase in the atom number, as seen when $N_f=90$,
results in an insulating region where $n_i^{uc}=1.5$.

A brief comment on the finite-size effect is in order here. 
Since we are dealing with the rather small system with 60 
fermions, the resulting metallic regions are very narrow, 
and the density profile changes very sharply there. This 
is due to the finite-size effect of our small system. If 
we consider a system with larger number of fermions, the 
metallic regions become wider and the density profile 
changes more smoothly.
\par
\subsection{Density fluctuations}

In order to characterize the above insulating regions,
 we further investigate local density fluctuations 
in our superlattice system. To this end, we 
introduce the variance of local density fluctuations, $\Delta_i$, 
at each site as
\begin{equation}
	\Delta_i \equiv \langle n_i^2 \rangle - 
	\langle n_i \rangle^2.
\label{33}
\end{equation}
In ordinary one-dimensional optical lattices, it is known that 
$\Delta_i$ is suppressed and
forms a plateau in the insulating region, while it has 
larger values in the metallic region \cite{Rigol,Rigol2}.
This quantity, therefore, can describe the way in which insulator-metal
crossover occurs when the on-site repulsion increases. 

\begin{figure*}[t]
\begin{center}
\includegraphics[width=14cm ]{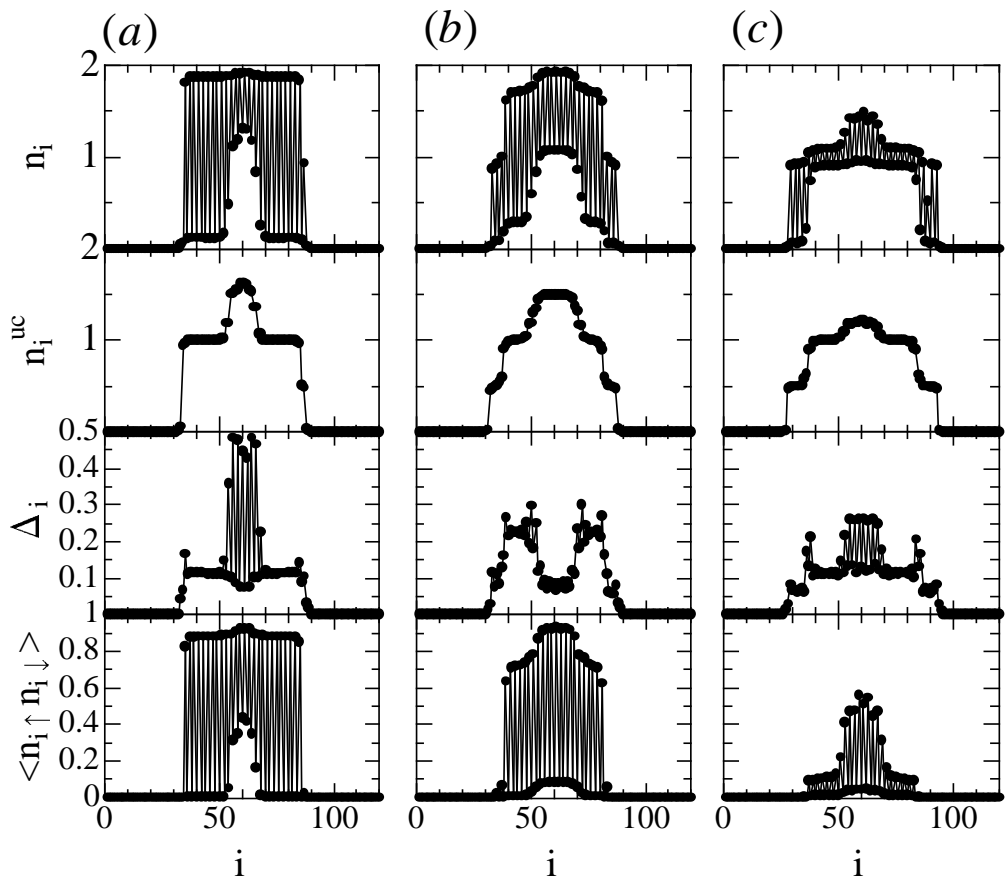}
\caption{
Plots of the local density $n_i$, the average density
$n_i^{uc}$, the variance $\Delta_i$
and the double-occupation probability 
$\langle n_{i \uparrow} n_{i \downarrow} \rangle$ for 
(a) $U/t=0.0$, (b) $3.0$, and (c) $9.0$.
The other parameters are fixed as $N_f=60$,
$N_s=120$, $V_c/t=0.01$, and $V_{SL}/t=5.0$.
}
\label{fig:many}
\end{center}
\end{figure*}

Figure\ \ref{fig:many} shows  
the local density $n_i$,  the average density
$n_i^{uc}$,  the 
variance $\Delta_i$  and the double-occupation probability
$\langle n_{i \uparrow} n_{i \downarrow} \rangle$
for three different choices of 
$U/t$ with the other parameters unchanged.

It is convenient to discuss characteristic properties by
specifying each region in terms of the average local 
density $n_i^{uc}$.
Let us first observe the insulating region characterized by $n_i^{uc}=1.0$,
which is located around the site $i=45$ (or equivalently $i=75$)
in all three cases in  Fig.\ \ref{fig:many}.
For $U/t=0.0$, we can see the sawtoothed oscillation both in 
$n_i$ and $\langle n_{i \uparrow} n_{i \downarrow} \rangle$
since the system is in the band insulator where 
almost doubly-occupied sites and vacant sites are realized alternately.
Note that even in this case, 
 the variance $\Delta_i$ forms a plateau 
without such alternations, implying that density fluctuations 
occur locally only via a process between two adjacent sites.
As the interaction strength $U$ increases,  $n_i$ 
($\langle n_{i \uparrow} n_{i \downarrow} \rangle$) continuously 
approaches $n_i \rightarrow 1$ 
($\langle n_{i \uparrow} n_{i \downarrow} \rangle \rightarrow 0$).
On the other hand, $\Delta_i$ is once enhanced
 ($U/t=3.0$, Fig.\ \ref{fig:many}(b)), and then
suppressed ($U/t=9.0$, Fig.\ \ref{fig:many}(c)).
Therefore, we naturally expect that there may be  
another insulating state, which is characterized by an enhanced  
$\Delta_i$, 
between  the band insulating
 state (small $U$) and the Mott insulating state (large $U$).
As mentioned above, in a homogeneous system without confinement, 
 at half filling (corresponding to $n_i^{uc} = 1.0$), the BCDW
intermediate  phase appears 
between the band-insulating phase and the Mott 
insulating phase \cite{Fabrizio,b12,b16,b17,b18,b19,b20,Manmana, Otsuka}.
In the BCDW state, there is 
 rather large hopping within a coupled dimer, which gives
rise to enhanced local density fluctuations. Therefore, we 
can distinguish the BCDW state from the other insulating states 
in terms of 
the variance $\Delta_i$.  Summarizing the above results,
we can say  that 
the region with   $n_i^{uc}=1.0$ is either in  the band insulating
state, the BCDW state or the Mott insulating state depending on the 
strength of the on-site repulsion. These states are smoothly connected 
to each other with the increase in $U$ (crossover behavior) 
in contrast to the 
 phase transitions realized in the homogeneous case.
 
We next focus on the region around the site $i=35$, where the average 
density  $n_i^{uc}=1.0$
at $U/t=0.0$ is changed to  $n_i^{uc}=0.5$ at $U/t=3.0$ and $ 9.0$.  In this case, 
the local density $n_i$ is reduced from 2 to 1 at each odd site with 
the increase  in $U$, and
$\Delta_i$ and $\langle n_{i \uparrow} n_{i \downarrow} \rangle$
decrease monotonically to zero
(even sites are always empty). Namely,  the 
system is smoothly driven from the band insulator to the Mott insulator.
Finally, we look at the region around $i=60$, where $n_i^{uc}$ is slightly
larger than 1.5 at $U/t=0.0$ without plateau structures, which is characteristic 
of the metallic region. In fact, $\Delta_i$ exhibits strong even-odd 
 dependence on the site index. At $U/t=3.0$, we see the plateau formation
at $n_i^{uc}=1.5$, where  
$n_i \sim 1.0$ and 
$\langle n_{i \uparrow} n_{i \downarrow} \rangle \sim 0$
 for even sites.
Therefore, the region enters the insulating region with 
Mott (odd sites) and band-insulating (even sites) characteristics.  
A further increase in 
$U$ again drives the Mott insulating 
state to the metallic state characterized by $1.0< n_i^{uc} <1.5$.


We have repeated similar calculations for various choices 
of the model parameters. Among them we have presented 
only the three selected cases above, since they 
capture most of the essential properties realized 
in our optical superlattice with 2-site periodicity.

\par
\subsection{Spin correlations}

\begin{figure}[t]
\includegraphics[width=0.7\linewidth]{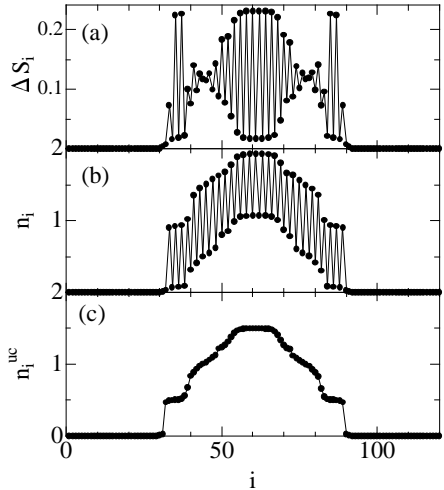}
\caption{Plots of (a) the variance of local spin fluctuations
 $\Delta S_i $
together with (b) $ n_i $ and (c) 
 $n_i^{uc}$. 
The data are obtained for the system with  $V_{SL}/t=5.0$, $U/t=5.0$.
 The other parameters are fixed as $N_s=120$, $N_f=60$, $V_c/t=0.01$.
}
\label{fig:ss}
\vspace{6pt}
\end{figure}

\begin{figure}[t]
\includegraphics[width=0.7\linewidth]{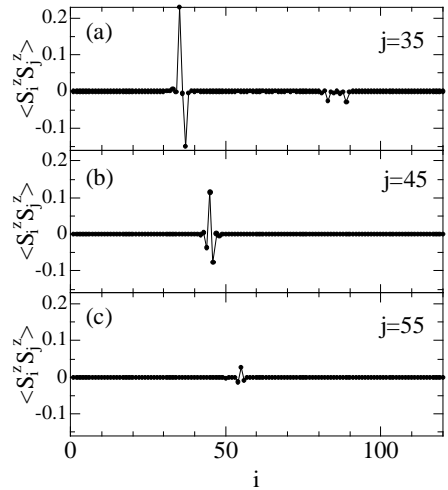}
\caption{Plots of the 
spin correlation function between the $i$th and $j$th sites,
$\langle S_i^z S_j^z \rangle$, where the site $j$ is fixed as
(a)$j=35$, (b)45 and (c)55.
The data are obtained for the system with  $V_{SL}/t=5.0$, $U/t=5.0$.
 The other parameters are fixed as $N_s=120$, $N_f=60$, $V_c/t=0.01$.
}
\label{fig:ss-1}
\end{figure}

Finally, we discuss local and spatially-extended
spin correlations in the system.
In Fig.\ \ref{fig:ss}(a)-(c), we show the computed results for 
the variance of local 
spin fluctuations $\Delta S_i = \langle (S_i^z)^2 \rangle$ 
together with the 
density profile $ n_i$ and the average profile
$n_i^{uc}$ for the system with $V_{SL}/t=5.0$ and $U/t=5.0$.
Also shown in Fig.\ \ref{fig:ss-1}(a)-(c) is
 the spin correlation function $\langle S_i^z S_j^z \rangle$ 
between the $i$th and $j$th  sites with the latter fixed  as 
three typical positions, $j=35$, 45, and 55.
\par
As clearly seen from Fig.\ \ref{fig:ss}, the local moment is well developed 
($ \Delta S_i  > 0.2$) when 
a given site is located in  the Mott insulating region, while 
it is  suppressed ($ \Delta S_i \sim 0$)
in the band-insulating region
as well as in the empty region.  These characteristic properties are
in accordance with what we naively expect from the homogeneous case
without harmonic confinement.
Let us now look briefly at the spin correlation function.
\par
In Fig.\ \ref{fig:ss-1}(a), the $i$-dependence of the  spin correlation 
$\langle S_i^z S_j^z \rangle$ is shown where
the site $j=35$ is fixed in the insulating region 
with $n_i^{uc}=0.5$. It is seen that the spin correlation
exhibits an antiferromagnetic nature within the same region, and 
quickly decays when the site $i$ enters the metallic region.
Interestingly, the spin correlation 
increases slightly again around $i=85$ belonging to
another insulating region with $n_i^{uc}=0.5$.
 This unusual behavior is
discussed below in detail by using a more significant example.
On the other hand, for $j=45$ in the insulating region with  
 $n_i^{uc}=1.0$,
we cannot see the long-range spin correlation as in the 
$j=35$ case, since this Mott insulating region is not 
well stabilized as regards the choice of parameters.
For $j=55$ in  Fig.\ \ref{fig:ss-1}(c), the spin correlation shows 
typical behavior inherent in the metallic state: a small 
spin correlation amplitude, which immediately decreases with distance.
\par
\begin{figure*}[t]
\begin{minipage}{.48\linewidth}
\includegraphics[width=0.7\linewidth]{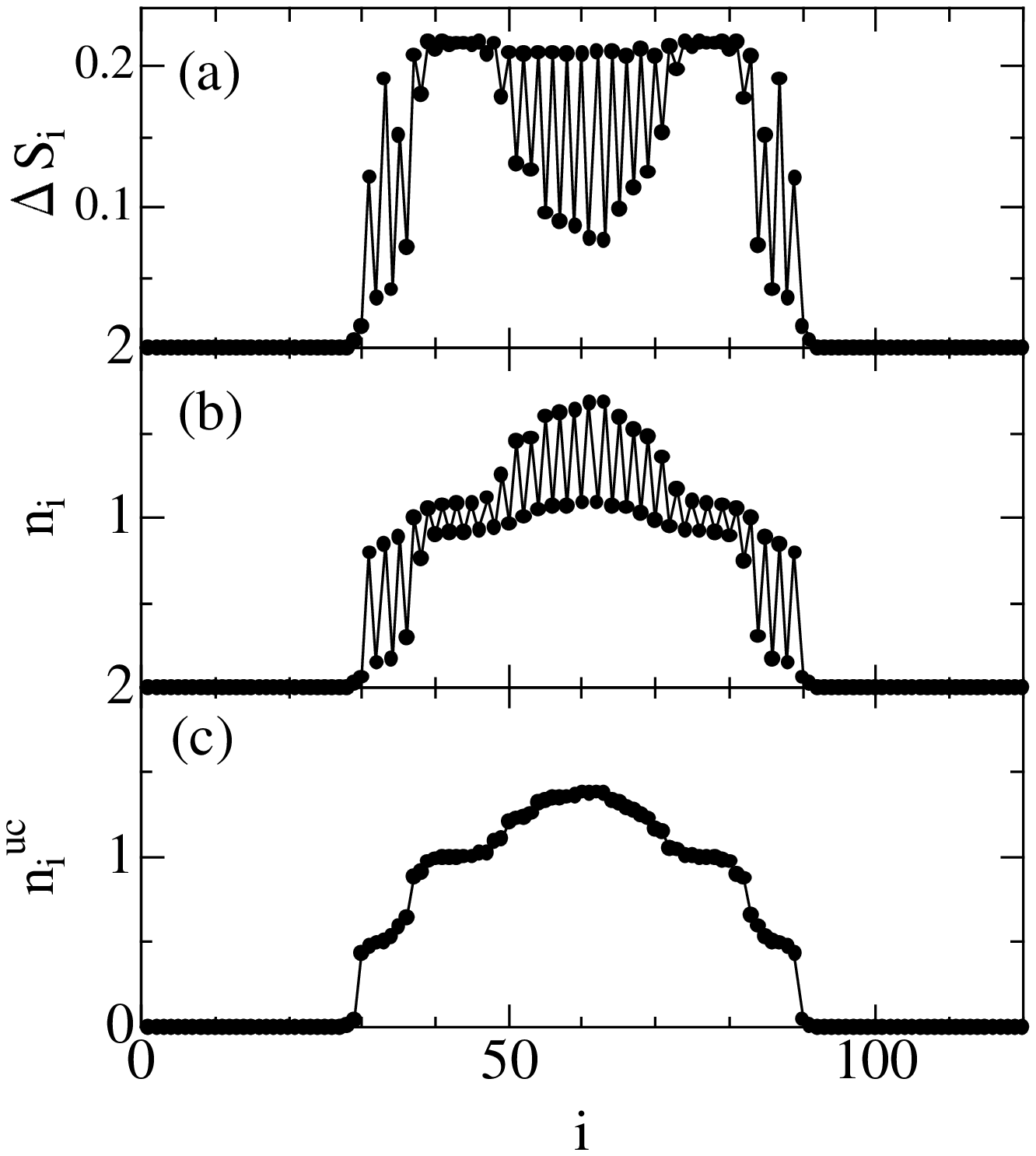}
\caption{Similar plots to those in Fig.\ref{fig:ss} for the other
choice of  $V_{SL}/t=3.0$ and $U/t=7.0$.
}
\label{fig:ss2}
\end{minipage}
\hspace{12pt}
\begin{minipage}{.48\linewidth}
\includegraphics[width=0.7\linewidth]{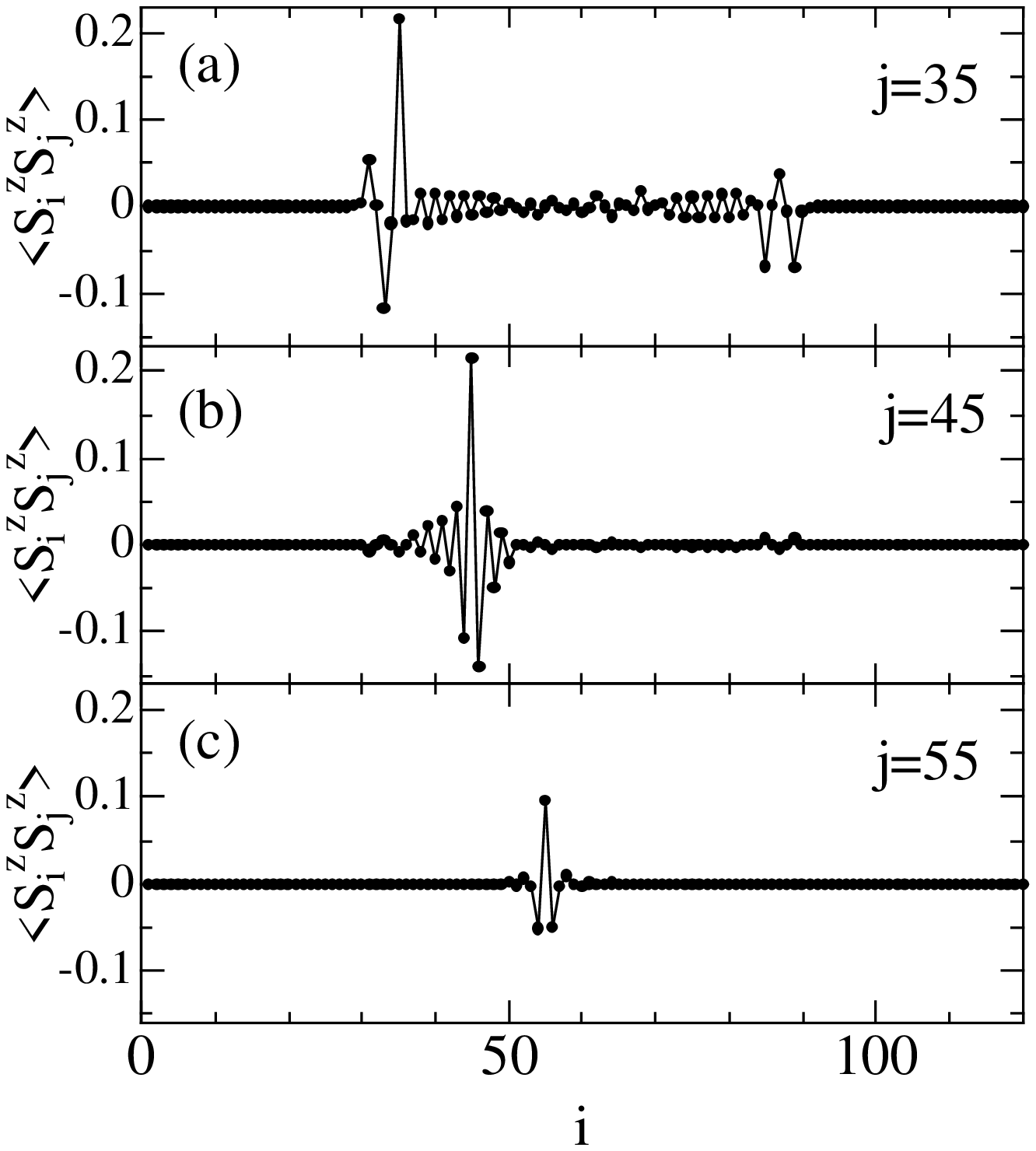}
\caption{Similar plots to those in Fig.\ref{fig:ss-1} for the other
choice of  $V_{SL}/t=3.0$ and $U/t=7.0$.
}
\label{fig:ss2-1}
\end{minipage}
\end{figure*}

In  Figs.\ \ref{fig:ss2} and \ref{fig:ss2-1}, we show the results 
obtained when we chose 
a slightly larger interaction strength $U$.  In this case also, the 
spin correlation shows similar behavior to that in Fig.\ \ref{fig:ss-1}.
There are two points to be mentioned here.
Since in this case the local moment is well developed in the Mott 
insulating region, as seen from Fig.\ \ref{fig:ss2}(a), 
 the spin correlation shows more significant 
 antiferromagnetic correlations. 
Even for site $j$ belonging to the metallic region, we can observe
 slightly enhanced antiferromagnetic correlations (Fig.\ \ref{fig:ss2-1}(c)).
Another important point to be noticed is that the spin correlations in 
Fig.\ \ref{fig:ss2-1}(a) are strongly enhanced around $i=85$, although 
they are suppressed in the region of $i <85$ (c.f. Fig.\ \ref{fig:ss-1}(a)).
This characteristic behavior is a new feature that has 
not been observed in  homogeneous 
systems without confinement. This is closely related to the formation
of several distinct regions. In the Mott insulating region, 
 a finite number of local spins are well developed, which are 
 sensitive to effective magnetic fields.  Such spins belonging to the
different Mott regions can 
have enhanced correlations even if they are separated by 
the metallic phases.  This is why we have encountered 
enhanced spin correlations between $j=35$ and $i=85$ 
in Fig.\ \ref{fig:ss2-1}(a).
It should be noted that this effect is particularly significant if there is an 
odd number of spins in the Mott region.  For example, 
in Figs.\ \ref{fig:ss-1}(a) and \ref{fig:ss2-1}(a), 3 sites around $j=35$
and $i=85$ belong to the well-defined Mott regions, which indeed 
exhibit enhanced spin correlations.
\par
Therefore, the above phenomenon is related to the finite-size effect of 
the Mott region, so that such behavior should be somewhat obscured 
when the Mott regions become large.
It is thus interesting to know whether such behavior can really be observed 
by future experiments in the optical lattices.
\par 
\section{Summary}
We have studied the characteristic properties of fermions 
trapped in a one-dimensional optical superlattice with 2-site periodicity.
We found that plateau regions appear 
in the profiles of the average number of atoms in the unit cell 
$n_i^{uc}$=0.5, 1.0, and 1.5. 
In contrast to the ordinary Hubbard model in the optical lattice, 
new insulating regions appear with 
$n_i^{uc}=$0.5 and 1.5 caused by the superlattice potentials.
Furthermore, it has been shown by the variance and double-occupation 
probability results that three different insulating regions of
band-, BCDW- and Mott-type emerge in the insulating region with 
$n_i^{uc}=1.0$, which are smoothly connected 
to each other via a crossover
as the strength of the on-site repulsion $U$ is altered.
\par
We have also found that it is possible to enhance spin correlations 
between different Mott regions, even if 
they are once suppressed in the intermediate 
metallic region. This interesting phenomenon is
 due to the finite-size effect of the insulating 
regions, which are naturally formed by harmonic confinement.
Specifically, it occurs more significantly 
when an effective number of sites in the Mott insulating region
is odd.
\par
Since systematic experimental studies on the one-dimensional optical 
superlattice may be soon within our reach, 
we hope that the characteristic properties discussed
for cold fermions in this paper will be observed in the near future.
\par
\begin{acknowledgments}
The numerical computations were carried out at the Supercomputer Center, 
the Institute for Solid State Physics, University of Tokyo. 
This work was supported by Grant-in-Aid for Scientific Research on 
Priority Areas (Grant No. 18043017 ) from The Ministry of Education, Culture,
Sports, Science and Technology of Japan.
\end{acknowledgments}

%

\end{document}